\documentclass[11pt]{article}
\usepackage{graphicx}
\usepackage{amsfonts}

\usepackage{amsmath}
\usepackage{amsthm}
\usepackage{amssymb}

\def\be{\begin{equation}}
\def\ee{\end{equation}}
\def\bea{\begin{eqnarray}}
\def\eea{\end{eqnarray}}

\def\1{\'{\i}}                           

\def\k{\omega}
\def\>#1{{\bf #1}}                 

\def\ro{\rho}

 \def\z{z}

 \def\adsw{AdS$_\omega$\ }

\def\co{\Delta}

\def\back{\!\!\!\!\!\!\!}            
\def\mback{\!\!\!\!}

\newcommand{\nn}{\nonumber}

\begin{document}

 \
 
\bigskip

\bigskip

  \begin{center}

 \noindent
  {\Large{\bf{On quantum deformations of (anti-)de Sitter algebras \\[6pt]in (2+1) dimensions }}}

\bigskip

{\sc {\'Angel Ballesteros$^1$, Francisco J. Herranz$^1$ and
Fabio Musso$^2$}}

\bigskip

 \noindent
 {$^1$ Departamento de F\1sica, Universidad de Burgos, 
09001 Burgos, Spain}

 \noindent
 {$^2$  Dipartimento di Fisica `Edoardo Amaldi', Universit\'a degli Studi di Roma Tre, 00146 Roma, Italy
}
 
E-mail: {angelb@ubu.es, fjherranz@ubu.es,  musso@fis.uniroma3.it}

\end{center}

\bigskip

\bigskip

\begin{abstract}  
Quantum deformations of (anti-)de Sitter  (A)dS algebras in (2+1) dimensions are revisited, and several features of these quantum structures are reviewed. In particular, the classification problem of (2+1) (A)dS Lie bialgebras is presented and the associated noncommutative quantum (A)dS spaces are also analysed. Moreover, the flat limit (or vanishing cosmological constant) of all these structures leading to (2+1) quantum Poincar\'e algebras and groups is simultaneously given by considering the cosmological constant as an explicit Lie algebra parameter in the (A)dS  algebras.
By making use of this classification, a three-parameter generalization of the $\kappa$-deformation for the (2+1) (A)dS algebras and quantum spacetimes is given. Finally, the same problem is studied in (3+1) dimensions, where a two-parameter generalization of the $\kappa$-(A)dS deformation that preserves the space isotropy is found.
\end{abstract}

\newpage


\section{Introduction}

Quantum groups and algebras were introduced in the eighties as Hopf algebra deformations of Lie groups and algebras, respectively (see~\cite{KR,Dri87, Jimbo, Takh, FRT, CP, majid} and references therein). Since then, the construction of quantum deformations of the kinematical groups of spacetimes opened the possibility of introducing consistent proposals of the mathematical descriptions of the `quantum' spacetime that is commonly assumed to arise in quantum gravity theories when the Planck energy is approached (see~\cite{Garay}). Essentially, this idea has been worked out so far in the literature under two dual approaches:
\begin{itemize}

\item {\em Quantum kinematical algebras}: they provide quantum deformations of the symmetry algebras of spacetimes (see, for instance,~\cite{kappaP,BHOS, beyond}) in which the `quantum' deformation parameter would be a second invariant scale related with the Planck energy (or length) and in which $q$-deformed Casimir operators generate the kind of modified dispersion relations that are expected to arise in the Planck regime. This viewpoint gave rise to the so-called `double special relativity'   theories~\cite{DSR,
MagueijoSmolin,Kowalski-Glikman:2002we,KowalskiFS,symmetry}.

\item {\em Quantum kinematical groups}: they are the Hopf algebra dual of the previous quantum algebras~\cite{Takh,FRT, Woronowicz} and provide a self-consistent mathematical foundation of the so-called noncommutative spacetimes, in which the noncommutativity of the spacetime coordinates is generated by a non-vanishing quantum deformation parameter that should describe Planck scale effects. As a consequence, uncertainty relations between these noncommuting coordinate operators arise naturally, and the intrinsic Planck scale `fuzziness', `discretization' or `quantization' of the spacetime itself can be mathematically described (see~\cite{Maggiore,FredenCMP,CKNT}).

\end{itemize}

At this point it is important to stress that for a given Lie group or algebra there exist (many) different Hopf algebra deformations. In fact, quantum algebras are in correspondence with Lie bialgebra structures (and quantum groups with Poisson--Lie structures), and the explicit classification problem for Lie bialgebras (or, equivalently, for Poisson--Lie groups) is by no means a simple problem from a computational viewpoint. Therefore, it seems natural that some criteria should be taken into account in order to select the type of quantum deformation that could be suitable in the abovementioned specific physical settings. 

In the particular case of $(2+1)$ quantum gravity, it was (heuristically) stated in~\cite{amel} that the perturbations of the vacuum state of a Chern--Simons   quantum gravity theory with cosmological constant $\Lambda$ are invariant under transformations that close a  certain quantum deformation of the (A)dS algebra. In fact, the low energy regime/zero-curvature limit of this algebra was found to be the known $\kappa$-Poincar\'e quantum algebra~\cite{kappaP,kMas,kMR,kZakr}.

Also, it is well known that Poisson--Lie (PL)  structures on the isometry groups of (2+1) spaces with constant curvature seem to play a relevant role as phase spaces in Chern--Simons   theory. Moreover, in the Lorentzian case, these PL structures (given by certain classical $r$-matrices that have to be `compatible' with the Chern-Simons pairings) are just the classical counterparts of certain quantum deformations of (A)dS and Poincar\'e groups in (2+1) dimensions (see~\cite{AMII,FR,cm1,cm2} and references therein). These PL groups are in one-to-one correspondence with certain three-dimensional   Lie bialgebras and their associated Drinfel'd doubles, in such a way that an explicit connection between (2+1) Chern-Simons symmetries and Drinfel'd doubles coming from  three-dimensional   quantum deformations has been recently proposed~\cite{BHMplb, BHMCQG}. In this setting the (2+1) (A)dS and Poincar\'e $r$-matrices associated to such Drinfel'd doubles are not the ones defining $\kappa$-deformations and, therefore, quantum groups corresponding to these (multiparametric) $r$-matrices seem to be physically interesting to construct and analyse. 


The aim of this paper is to address the classification problem for the  quantum deformations of the (2+1) (A)dS and Poincar\'e algebras by considering the three Lie algebras as particular cases of a one-parametric Lie algebra, the (2+1) \adsw algebra, in which $\k=-\Lambda$. In this way, the $\k\to 0$ limit will provide the Poincar\'e Lie bialgebra classification, and the sign of $\k$ will distinguish between the AdS ($\k>0$) and dS ($\k<0$) cases. Moreover, we will show that the type of approach here presented makes feasible to tackle the same problem in (3+1) dimensions.

The paper is organized as follows. In the next section we will briefly recall the basics of quantum deformations and their classification in terms of their associated Lie bialgebras. In section 3 we will review the known results concerning the $\kappa$-deformation of the (2+1) \adsw algebra, including the full PL structure that provides the semiclassical (Poisson) counterpart of the noncommutative (2+1) \adsw spacetime, where the nonvanishing cosmological constant $\Lambda$ generates nonlinear commutation relations among the quantum coordinates. In section 4 the most generic Lie bialgebra for the (2+1) \adsw algebra is  computed, and this result gives a clear idea of the plethora of different quantum deformations that there exist. The possible generalizations of the $\kappa$-deformation are identified in section 5 by imposing   on the generic deformation that the rotation $J$ and time translation $P_0$ generators have to be primitive ones, a constraint derived from dimensionality arguments. In this way we find that two more parameters can be added to the $\kappa$-deformation, and we explicitly describe the corresponding first-order noncommutative spacetimes, which are generalizations of the $\kappa$-Minkowski spacetime. The same approach is used in section 6 in order to study the possible generalizations of $\kappa$-deformations in (3+1) dimensions. In this case we find two disjoint families of two-parametric deformations that preserve the primitive nature of the generators $P_0$ and $J_3$. Finally, we demonstrate that one of these generalized $\kappa$-deformations preserve space isotropy, since under some appropriate automorphism the symmetric role of the three rotation generators can be manifestly shown.






\section{Quantum deformations and Lie bialgebras}

\subsection{Quantum algebras}

Quantum algebras 
are Hopf algebra deformations of universal enveloping algebras. This means that we consider the algebra $U_z(g)$ of formal power series in the deformation parameter $z$ and coefficients in the universal enveloping algebra $U(g)$ of a given Lie algebra $g$, and we endow it with a Hopf algebra structure by finding an algebra homomorphism called the coproduct map
$$
\Delta_U: U_z(g)\longrightarrow U_z(g)\otimes U_z(g) ,
\label{cou}
$$
as well as its associated counit $\epsilon_U$ and antipode $\gamma_U$ mappings. 

Any quantum universal enveloping algebra can be thought of as a Hopf algebra deformation of $U(g)$ `in the direction' of a certain Lie bialgebra $(g,\delta)$. Such a Lie bialgebra provides the first-order deformation (in $z$) of the coproduct
$$
\Delta_U
=\Delta_0 + \delta+ o[z^2] ,
\label{powerco}
$$
where $\Delta_0(X)=X \otimes 1+1\otimes X $ is the primitive (nondeformed) coproduct and  $\delta(X)$, the so-called  cocommutator,  will be linear in $z$.
Therefore, a precise Lie bialgebra structure $(g,\delta)$ will be associated to each possible quantum coproduct $\Delta_U$, and the equivalence classes (under automorphisms) of Lie bialgebra structures on a given Lie algebra $g$ will provide the chart of all possible quantum deformations of $U(g)$.

Complementarily, quantum groups $(A,\Delta_A)$ are just noncommutative algebras of functions defined as the dual Hopf algebras  to the quantum algebras $(U_z(g),\Delta_U)$. More explicitly, if we denote by $m_A$ and $m_U$ the noncommutative products in $A$ and $U_z(g)$, respectively, the duality between the Hopf algebras  $(A,m_A,\co_A)$ and
$(U_z(g),m_U,\co_U)$ is established through the existence of a
canonical pairing
$\langle
\, , \, \rangle: A\times U_z(g)\rightarrow \mathbb C$ between them such that
\bea
&& \langle m_A(f\otimes g), a \rangle=\langle f\otimes g ,
\co_U(a)\rangle,\label{pairing1}
\\
&& \langle \co_A(f) , a\otimes b \rangle= \langle f,
m_U(a\otimes b)\rangle,
\label{pairing2}
\eea
where $a,b\in U_z(g)$; $f,g\in A$, and 
$
\langle f\otimes g , a\otimes b \rangle=\langle f , a
\rangle\,\langle g , b \rangle.
$

It is important to stress that the duality relation (\ref{pairing1}) implies that the product in $A$ is defined by the coproduct in $U_z(g)$, and conversely relation (\ref{pairing2}) implies that the coproduct in $A$ is given by the product in $U_z(g)$. Since the first-order deformation of the coproduct $\Delta_U$ is defined by the Lie bialgebra map $\delta$, the first-order noncommutativity for the quantum group $A$ will be given by (the dual of) $\delta$.  Therefore, the Lie bialgebra structure associated to a given quantum algebra provides immediately the first-order information about the noncommutative algebra of quantum group coordinates. Moreover, the Lie bialgebra $(g,\delta$) is in one-to-one correspondence with the PL structure whose quantization provides the full quantum group $(A,\Delta_A)$. As a consequence, the classification of all possible Lie bialgebra structures for $g$ constitute the first and most relevant step for the analysis and explicit construction of its quantum algebra deformations.


\subsection{Lie bialgebras}

Let us summarize all the basic facts about Lie bialgebras that will be needed in the sequel.
A Lie bialgebra $(g,\delta)$ is Lie algebra $g$ with structure tensor $c^k_{ij}$
\be
[X_i,X_j]=c^k_{ij}X_k ,
\label{liealg}
\ee
that is endowed with a skewsymmetric cocommutator map
$$
\delta:{g}\to {g}\otimes {g}
$$
fulfilling the two following conditions:
\begin{itemize}
\item i) $\delta$ is a 1-cocycle, {\em  i.e.},
$$
\delta([X,Y])=[\delta(X),\,  Y\otimes 1+ 1\otimes Y] + 
[ X\otimes 1+1\otimes X,\, \delta(Y)] ,\qquad \forall \,X,Y\in
{g}.
\label{1cocycle}
$$
\item ii) The dual map $\delta^\ast:{g}^\ast\otimes {g}^\ast \to
{g}^\ast$ is a Lie bracket on ${g}^\ast$.
\end{itemize}
Therefore any cocommutator $\delta$ will be of the form
\be
\delta(X_i)=f^{jk}_i\,X_j\wedge X_k \, ,
\label{precoco}
\ee
where $f^{jk}_i$ is the structure tensor of the dual Lie algebra $g^\ast$ defined by
\be
[\hat\xi^j,\hat\xi^k]=f^{jk}_i\,\hat\xi^i \, ,
\label{dualL}
\ee
where $\langle  \hat\xi^j,X_k \rangle=\delta_k^j$. Therefore, a Lie bialgebra is a pair of `matched' Lie algebras of the same dimension, since the 1-cocycle condition implies the following compatibility equations
\be
f^{ab}_k c^k_{ij} = f^{ak}_i c^b_{kj}+f^{kb}_i c^a_{kj}
+f^{ak}_j c^b_{ik} +f^{kb}_j c^a_{ik}. 
\label{compatfc}
\ee

As it could be expected, for some Lie bialgebras the 1-cocycle $\delta$ is a coboundary
\be
\delta(X)=[ X \otimes 1+1\otimes X ,\,  r],\qquad 
\forall\,X\in {g} ,
\label{cocom}
\ee
where $r$ is a skewsymmetric element of
${g}\otimes {g}$ (the classical $r$-matrix)
$$
r=r^{ab}\,X_a \wedge X_b\, ,
$$
that has
to be a solution of the modified classical Yang--Baxter equation (mCYBE)
\be
[X\otimes 1\otimes 1 + 1\otimes X\otimes 1 +
1\otimes 1\otimes X,[[r,r]]\, ]=0, \qquad \forall X\in {g},
\label{mCYBE}
\ee
in which the Schouten bracket $[[r,r]]$ is defined as
\be
[[r,r]]:=[r_{12},r_{13}]+ [r_{12},r_{23}]+ [r_{13},r_{23}] ,
\label{Schouten}
\ee
  where $r_{12}=r^{ab}\,X_a \otimes X_b\otimes 1, r_{13}=r^{ab}\,X_a \otimes 1\otimes X_b, r_{23}=r^{ab}\,1 \otimes X_a\otimes X_b$. Recall that $[[r,r]]=0$ is just the classical Yang--Baxter equation (CYBE).

For semisimple Lie algebras all Lie bialgebra structures are coboundaries, and that is also the case for the Poincar\'e algebra in (2+1) and (3+1) dimensions. On the contrary, for solvable and nilpotent Lie algebras many of their Lie bialgebra structures are non-coboundaries (see~\cite{gomez,BBMpl} and references therein).

\subsection{Classification of Lie bialgebra structures on $g$}

With the previous definitions in mind, the algorithm for the explicit computation and classification of all the possible Lie bialgebra structures for a given Lie algebra $g$ (\ref{liealg}) in the basis $X_i$ can be sketched as follows:
\begin{itemize}

\item The most generic (pre)-cocommutator map is given  in the form (\ref{precoco}).

\item The 1-cocycle condition (\ref{compatfc}) is imposed onto $f^{jk}_i$. 

\item Further, the (quadratic) condition that the dual map 
$
\delta^\ast:{g}^\ast\otimes {g}^\ast \to
{g}^\ast
$ is a Lie bracket on ${g}^\ast$ is imposed. Note that this constraint is nothing but the Jacobi identity for $f^{jk}_i$.

\item Finally, the solutions for $f^{jk}_i$ can be classified into equivalence classes under automorphisms of the Lie algebra $g$.

\end{itemize}

Note that since Lie bialgebra structures for (2+1) (A)dS and Poincar\'e algebras are known to be always coboundaries, this procedure is completely equivalent to the classification of constant classical $r$-matrices on these algebras.  

\section{The $\kappa$-deformation of the \adsw algebra}

In this section we exemplify the theory of quantum deformations by recalling the main results concerning the so-called $\kappa$-deformation of the (2+1) (anti-)de Sitter and Poincar\'e algebras. We will use a unified approach to the three Lie algebras by making use of the six-dimensional Lie algebra \adsw which is given in terms of the generators
$\{J,P_0,P_i, K_i\}$ $(i=1,2)$ (rotation, time translation, space translations and boosts) as
\be
\begin{array}{lll} 
[J,P_i]=   \epsilon_{ij}P_j , &\qquad
[J,K_i]=   \epsilon_{ij}K_j , &\qquad  [J,P_0]= 0  , \\[2pt]
[P_i,K_j]=-\delta_{ij}P_0 ,&\qquad [P_0,K_i]=-P_i ,&\qquad
[K_1,K_2]= -J   , \\[2pt]
[P_0,P_i]=\k K_i ,&\qquad [P_1,P_2]= -\k J  ,
\end{array}
\label{ea}\nonumber
\ee
where $i,j=1,2$, $\epsilon_{12}=1$ and  the parameter $\k$   is just the constant sectional curvature of the corresponding classical spacetime (so this is related to the cosmological constant through $\k=-\Lambda$). Therefore, these Lie brackets encompass:   
\begin{itemize}

\item The AdS Lie   algebra, $so(2,2)$, when  $\k=+1/R^2>0$.

\item The dS Lie algebra, $so(3,1)$, when $\k=-1/R^2<0$.

\item The Poincar\'e Lie  algebra, $iso(2,1)$, when $\k=0$. Note that this case corresponds to the flat  contraction obtained via the limit of the universe radius  $R\to \infty$  connecting   $so(2,2)\to iso(2,1)\leftarrow so(3,1)$.

\end{itemize}

The two Casimir invariants of the \adsw algebra are given by
$$
{\cal C}=P_0^2-\>P^2+\k(J^2-\>K^2), \qquad
{\cal W}=-JP_0+K_1P_2-K_2P_1  ,
\label{bc}
$$
where $\cal C$ comes from the Killing--Cartan form and it is related to the energy of the particle, while $\cal W$ is the Pauli--Lubanski vector.

\subsection{The Lie bialgebra of the $\kappa$-deformation}

As we mentioned before, all Lie bialgebra structures for the (A)dS and Poincar\'e Lie algebras in (2+1) dimensions are coboundaries. This means that the Lie bialgebra is fully specified by a certain skewsymmetric solution of the mCYBE (or of the CYBE). In the case of the $\kappa$-deformation such a classical $r$-matrix reads
\be
r=z(K_1\wedge P_1+K_2\wedge P_2),
\label{ca}
\ee
where  the quantum deformation parameter $z$ is related to the usual Planck energy scale $\kappa$ through $z=1/\kappa$.
Once $r$ is fixed, the cocommutator $\delta$ providing the first-order deformation of the coproduct is given by the coboundary relation (\ref{cocom}), and turns out to be  
\be
\begin{array}{l} 
\delta(P_0)=0 ,  \qquad \delta(J)=0 , \\[2pt]
\delta(P_i)=z(P_i\wedge P_0-\k\epsilon_{ij} K_j\wedge J) , \\[2pt]
\delta(K_i)=z(K_i\wedge P_0+\epsilon_{ij} P_j\wedge J) . 
\end{array}
\label{cok}
\ee

A `quantum group local coordinate' $\hat y^i$ will be defined as dual to the generator
$Y_i$ through the Hopf algebra pairing  $\langle \hat y^i ,Y_j\rangle=    \delta_j^i$.
In particular,   we will denote    $\{ \hat\theta, \hat
x_\mu,  \hat\xi_i \} $   the noncommutative coordinates which are dual to the 
generators $\{J,P_\mu,K_i\}$ $(\mu=0,1,2)$, respectively. Now, if we make use of the dual Lie algebra (\ref{dualL}) induced by the first-order deformation (in $z$) of the coproduct given by $\delta$ (\ref{cok}), we get the first-order quantum group relations for the $\kappa$-deformation of the \adsw algebra, namely:
\be
\begin{array}{llll} 
[\hat x_0, \hat x_i]=-z \hat x_i  ,&\quad
[\hat\theta,\hat x_i]= z\epsilon_{ij}  \hat \xi_j ,&\quad
[\hat x_1, \hat x_2]=0,&\quad [\hat\theta,\hat x_0]=0 ,\\[2pt]
[\hat\theta,\hat \xi_i]=-z\k \epsilon_{ij}  \hat x_j ,&\quad
[\hat x_0, \hat \xi_i]=-z \hat \xi_i ,&\quad
[\hat \xi_1, \hat \xi_2]=0,&\quad [\hat x_i,\hat \xi_j]=0 .
\end{array}
\label{firstqg}\nonumber
\ee

We realize that in these relations the `quantum' time and space translation parameters close a non-Abelian subalgebra
\be
[\hat x_0, \hat x_i]=-z \hat x_i ,\qquad
[\hat x_1, \hat x_2]=0,
\qquad
i=1,2,
\label{kminkow}
\ee
that in the case of the quantum Poincar\'e group is known as the (2+1) $\kappa$-Minkowski noncommutative spacetime ${\bf M}_z^{2+1}$~\cite{kappaP,kMas,kMR,kZakr}, since no higher order corrections have to be incorporated when the full quantum Poincar\'e group is constructed. Note that commutativity is always recovered in the limit $z\to 0$, {\em i.e.}, when the quantum deformation vanishes. It is also worth stressing that the first-order relations (\ref{kminkow}) do not depend on the curvature $\k$, so the three first-order (A)dS and Minkowskian  noncommutative spacetimes coincide. As we shall see later, higher order corrections depending on $\k$ will appear when the full quantum (A)dS groups are considered.

\subsection{The   quantum $\kappa$-\adsw algebra}

The full (all orders in $z$) quantum universal enveloping algebra corresponding to the $\kappa$-defor\-mation of \adsw was constructed for the first time in~\cite{BHOS}. Its quantum coproduct reads (hereafter we will use the $\rho$ parameter defined as $\k=\ro^2$)
\be
\begin{array}{l} 
 \Delta(P_0)=P_0\otimes 1 + 1\otimes P_0,\qquad
\Delta(J)=J\otimes 1+ 1\otimes J ,\\[2pt]
 \Delta(P_i)=
P_i \otimes {\rm e}^{\frac z2  P_0 }\cosh(\frac z2 \rho J) + {\rm e}^{-\frac z2 P_0 }\cosh(\frac z2 \rho J)\otimes P_i\\[4pt]
\qquad\qquad-  \rho\,  \epsilon_{ij} K_j \otimes {\rm e}^{\frac z2 
P_0 }\sinh(\frac z2 \rho J)  +  \rho\, {\rm e}^{-\frac z2 P_0 }\sinh(\frac z2 \rho
J)\otimes \epsilon_{ij} K_j,\\[4pt]
\Delta(K_i)=
K_i \otimes {\rm e}^{\frac z2  P_0 }\cosh(\frac z2 \rho J) + {\rm e}^{-\frac z2 P_0 }\cosh(\frac z2 \rho J)\otimes K_i\\[4pt]
\qquad\qquad 
\displaystyle{ + \epsilon_{ij} P_j \otimes {\rm e}^{\frac z2 
P_0 }  \, \frac{ \sinh(\frac z2 \rho
J) }{\rho} } -   {\rm e}^{-\frac z2 P_0 }\, \frac{ \sinh(\frac z2 \rho
J) }{\rho}\otimes \epsilon_{ij} P_j,
\end{array}
\label{eh}\nonumber
\ee
the compatible set of deformed commutation rules are
\be
\begin{array}{l} 
[J,P_i]=   \epsilon_{ij}P_j  ,\qquad\ \,
[J,K_i]=   \epsilon_{ij}K_j  , \qquad\   [J,P_0]= 0   ,\\[4pt]
\displaystyle {[P_i,K_j]=-\delta_{ij}\frac{\sinh (zP_0)}{z}\,\cosh(z\rho J)}
, \qquad\, [P_0,K_i]=-P_i  , \\[2pt]
\displaystyle { [K_1,K_2]= -\cosh (zP_0)\, 
\frac{\sinh(z\rho J) }{z \rho} }  , \qquad\quad 
[P_0,P_i]=\k K_i   ,  \\[8pt] 
\displaystyle {  [P_1,P_2]= -\k \cosh (zP_0)  \, \frac{
\sinh(z\rho J) }{z \rho} }  ,
\end{array}
\label{ei}\nonumber
\ee
and the deformed Casimir operators have the form  
\be
\begin{array}{l} 
{\cal C}_z=4 \cos (z\rho) \left\{   {\displaystyle { \frac{\sinh^2(\frac z 2
P_0)}{z^2} }} \, \cosh^2\left(\frac z 2\rho J\right)  +
{\displaystyle {  \frac{\sinh^2(\frac z 2
\rho J)}{z^2}}}  \, \cosh^2\left(\frac z 2 P_0\right)
\right\}  
{\displaystyle {- \frac{\sin (z\rho)}{z\rho} \left(\>P^2+\k \>K^2 \right) }} ,\\[6pt]
\displaystyle {{\cal W}_z= -\cos (z\rho)\, \frac{\sinh(z\rho J)}{z\rho}\, 
\frac{\sinh(z P_0)}{z}+\frac{\sin (z\rho)}{z\rho} (K_1P_2-K_2P_1 )} .
\end{array}
\label{fc}\nonumber
\ee

It is worth mentioning that this is exactly the quantum (A)dS algebra proposed in~\cite{amel} as the symmetry algebra of the vacuum excitations in (2+1) quantum gravity. By construction, the limit $\k\to 0$ of this quantum algebra is always well defined and leads to the well known (2+1) $\kappa$-Poincar\'e algebra. Note also that the physical dimensions  of  the quantum deformation parameter $z$ are inherited from $P_0$, since $[z]=[P_0]^{-1}$. If $c=1$,  this means that $z$ can be interpreted as a fundamental length parameter, which in the usual double special relativity theories is considered to be of the order of the Planck length $l_p$. Algebraically, this link between $z$ and $P_0$ is directly related to the fact that $P_0$ and $J$  remain nondeformed (primitive) at the level of the coproduct, which allows the deformation to contain formal power series of these two generators (see~\cite{LBC} for a dimensional analysis of the deformation parameters from the viewpoint of contraction theory).

\subsection{The $\kappa$-\adsw Poisson--Lie group}

The full $\kappa$-\adsw quantum group could be obtained by computing the Hopf algebra dual  to the quantum $\kappa$-\adsw   algebra that we have just introduced. This requires a lengthy and cumbersome computation, but many features of this quantum group can be extracted from its semiclassical limit, {\em i.e.}, the PL  structure on the classical \adsw group that is in one-to-one correspondence with the Lie bialgebra that characterizes this quantum deformation. 

The construction of the $\kappa$-\adsw PL group was fully performed in~\cite{Rox}, where we refer to the interested reader for details.  In particular, under the appropriate parametrization of the group, it is found that the local coordinate functions corresponding to the translation parameters close the following PL  subalgebra
\be 
\begin{array}{l} 
\displaystyle{ \{x_0,x_1\} =-z\,\frac{\tanh\rho x_1}{\rho \cosh^2\!\rho x_2}, \qquad
\{x_0,x_2\} =-z\,\frac{\tanh\rho x_2}{\rho } ,\qquad 
\{x_1,x_2\} =0 } ,
\end{array}
\label{gc}\nonumber
\ee
whose quantization (in the usual `quantum-mechanical' sense) would give rise to the full noncommutative $\kappa$-\adsw spacetime. Since $\{x_1,x_2\} =0$, no ordering ambiguities appear in the quantization process, and the quantum (2+1) noncommutative \adsw spacetime
reads
\be 
\begin{array}{l} 
{\displaystyle{ [\hat x_0, \hat x_1] =-z\,\frac{\tanh\rho \hat x_1}{\rho
\cosh^2\rho
\hat x_2} = }} -z \hat x_1 + \frac 13 z\k \hat x_1^3+ z\k\hat x_1\hat
x_2^2 +o(\k^2) ,\\[10pt]
{\displaystyle{[\hat x_0,\hat x_2] =-z\,\frac{\tanh\rho \hat x_2}{\rho
}=}}   -z \hat x_2+\frac 13 z \k \hat x_2^3+o(\k^2) ,\qquad
 [\hat x_1,
\hat x_2] =0.  
\end{array} 
\label{ha}\nonumber
\ee

Therefore, the $\kappa$-Minkowski space ${\bf M}_z^{2+1}$ is the first-order  
noncommutative spacetime for all the \adsw quantum  groups. However, when the curvature is not zero ($\k\neq 0$) higher order contributions appear, and the quantum (A)dS spacetimes turn out to be nonlinear noncommutative algebras. Note also that, in any case, (see~\cite{Rox} for the full expressions) the quantum spacetime coordinates $\{\hat x_0, \hat x_1,\hat x_2\}$ close a subalgebra.

\section{The generic \adsw deformation}

The aim of the present section is to show that the $\kappa$-deformation is a very particular one-parametric choice amongst all the possible quantum deformations of the \adsw algebra. In the sequel we will present the chart of Lie bialgebra structures of the \adsw algebra, that can be obtained by taking into account that the Lie bialgebra structures of $so(2,2)$, $so(3,1)$ and the (2+1) Poincar\'e algebra are always coboundary structures.
Therefore, they come from classical $r$-matrices, and the most general form for a constant classical $r$-matrix on \adsw is
\begin{eqnarray}
r\!&\!=\!&\!a_1 J \wedge P_1 + a_2 J\wedge K_1 + a_3 P_0\wedge P_1 + a_4 P_0 \wedge K_1 +a_5 P_1\wedge K_1 +a_6 P_1\wedge K_2 \cr 
 &&+\ b_1 J\wedge P_2 + b_2 J\wedge K_2 + b_3 P_0\wedge P_2 + b_4 P_0\wedge K_2 +b_5 P_2\wedge K_2 + b_6 P_2\wedge K_1 \label{genericr}\\ 
 &&+\  c_1 J\wedge P_0 + c_2 K_1\wedge K_2 + c_3 P_1\wedge P_2 ,
\nonumber
\end{eqnarray}
where we have initially 15 possible `deformation parameters', that will have to fulfil the constraints coming form the mCYBE (\ref{mCYBE}). We recall that pure non-standard/twist \adsw deformations will be obtained when the Schouten bracket $[[r,r]]=0$  ({\em i.e.}, if their $r$-matrices are solutions of the CYBE).  

Explicitly, the mCYBE (\ref{mCYBE}) leads to the following nonlinear constraints involving the 15 (real) Lie bialgebra parameters and the curvature $\k$:
\begin{eqnarray}
&&(a_1 a_5 - a_4 a_6 + a_6 b_1 - a_1 b_5 + a_4 b_6 + b_1 b_6 + a_3 c_2 -\k  a_3
c_3)=0 , \nn\\
&&(a_2 a_3 - a_4 b_4 + b_1 b_4 + a_5 b_6 + b_5 b_6 + a_5 c_1 - b_5 c_1 - \k  a_3
b_3)=0, \nn\\
&&(-a_2 a_5 - a_6 b_2 + a_2 b_5 - b_2 b_6 + a_4 c_2 + \k a_3 a_6  - \k a_3 b_6  -
\k a_4 c_3 )=0, \nn\\
&&(-a_1 a_5 + a_4 a_6 - a_6 b_1 - a_5 b_4 + a_4 c_1 + b_1 c_1 + b_2 c_3 + \k a_3
c_3 )=0, \nn\\
&&(a_2 a_4 + a_2 b_1 - a_1 b_2 - a_5 c_2 + \k a_1 a_3  - \k a_5 c_3 )=\k  (a_1 a_3 +
a_4 b_3 + b_1 b_3 - a_3 b_4 - a_5 c_3 - b_5 c_3), \nn\\
&&(a_1 a_2 + b_1 b_2 + a_2 b_4 - a_6 c_2 - c_1 c_2 + \k a_3 b_1  - \k a_6 c_3  + \k c_1
c_3 )=0,\nn\\
&&(-a_2 a_5 - b_2 b_6 + b_2 c_1 + a_4 c_2 - b_1 c_2 + \k a_5 b_3  - \k a_3 b_6  -
\k a_3 c_1 )=0 ,
\nn\\&&(a_1 a_6 - a_5 b_1 - a_6 b_4 + b_1 b_5 + a_1 b_6 + b_4 b_6 + b_3 c_2 - \k b_3 c_3
)=0, \nn\\
&&(a_1 a_4 - a_5 a_6 + b_2 b_3 + a_4 b_4 - a_6 b_5 - a_5 c_1 + b_5 c_1 + \k a_3 b_3
)=0, \nn\\
&&(a_2 a_6 - a_5 b_2 + b_2 b_5 + a_2 b_6 - b_4 c_2 - \k a_6 b_3  + \k b_3 b_6  + \k b_4
c_3 ) =0, \nn \\
&&(-a_4 b_5 + b_1 b_5 + a_1 b_6 + b_4 b_6 + a_1 c_1 - b_4 c_1 + a_2 c_3 - \k b_3
c_3 )=0, \nn\\
&&(-a_2 b_1 + a_1 b_2 - b_2 b_4 + b_5 c_2 - \k b_1 b_3  +\k  b_5 c_3 )\nn\\
&&\qquad\qquad
=-\k (a_1 a_3 +
a_4 b_3 + b_1 b_3 - a_3 b_4 - a_5 c_3 - b_5 c_3), \nn\\
&&(a_1 a_2 - a_4 b_2 + b_1 b_2 + b_6 c_2 - c_1 c_2 - \k  a_1 b_3 + \k b_6 c_3  + \k c_1
c_3 )=0, \nn\\
&&(-a_2 a_6 - b_2 b_5 - a_2 c_1 + a_1 c_2 + b_4 c_2 + \k a_6 b_3  - \k a_3 b_5  -\k 
b_3 c_1 )=0,
\nn\\
&&(a_2a_4+b_2b_4-a_5c_2-b_5c_2+\k a_4b_3-\k a_3b_4)=\k  (a_1 a_3 + a_4 b_3 + b_1 b_3 - a_3 b_4 - a_5 c_3 - b_5
c_3), \nn\\ 
&&a_1^2+b_1^2-a_3b_2+a_2b_3-a_5b_5+a_6b_6-a_6c_1+b_6c_1-\k c_3^2   \nn\\
&&\qquad\qquad=
(a_4^2 - a_5^2 + a_2 b_3 - a_1 b_4 - a_6 b_6 + a_6 c_1 + b_6 c_1 - c_2 c_3
+\k a_3^2 ), \nn\\
&&( - a_2^2  -   b_2^2  +  c_2^2  -  \k a_4 b_1   +  \k a_1 b_4   +  \k a_5
b_5   -  \k a_6 b_6   -  \k a_6 c_1   +  \k b_6 c_1 )\nn\\
&&\qquad\qquad=-\k (a_4^2  -  a_5^2  + 
a_2 b_3  -  a_1 b_4  -  a_6 b_6  +  a_6 c_1  +  b_6 c_1  -  c_2 c_3 
+  \k a_3^2 ), \nn\\
&&(a_4 b_1 - a_3 b_2 + b_4^2 - b_5^2 - a_6 b_6 - a_6 c_1 - b_6 c_1 - c_2 c_3 + \k b_3^2 )\nn\\
&&\qquad\qquad=(a_4^2 - a_5^2 + a_2 b_3 - a_1 b_4 - a_6 b_6 + a_6 c_1 + b_6 c_1 - c_2 c_3 +\k a_3^2 ) . \label{constraints}
\end{eqnarray}
Therefore, the most generic cocommutator is given by (\ref{cocom}), namely:
\begin{eqnarray}
&&\back\mback  \delta(J) = a_1 J\wedge  P_2+ a_2 J\wedge  K_2 + a_3 P_0\wedge  P_2+ a_4 P_0\wedge  K_2 + (b_5 -a_5) (  K_1\wedge  P_2 - P_1\wedge  K_2) \nn\\
&&\back  +(a_6+b_6)( K_1\wedge  P_1 + P_2\wedge  K_2 ) + b_1 P_1\wedge  J + b_2K_1\wedge  J +  b_3 P_1\wedge  P_0 + b_4 K_1\wedge  P_0  ,\nn\\[0.1cm]
&&\back\mback  \delta(P_0) = \k  a_1  J\wedge K_1+ a_2 P_1\wedge J+ \k a_3  P_0\wedge K_1+a_4 P_1\wedge P_0  +(a_6 - b_6) ( P_2\wedge P_1+\k  K_1\wedge K_2 )    \nn\\
&& \back + \k b_1  J\wedge K_2 + b_2 P_2\wedge J  +\k b_3  P_0\wedge K_2+ b_4 P_2\wedge P_0+ (\k c_3 -c_2) (K_1\wedge P_2+ P_1\wedge K_2) ,\nn\\[0.1cm]
&&\back\mback  \delta (P_1)= (  b_6  + c_1 ) ( P_0 \wedge P_2 +\k  K_1\wedge J)+ \k   b_5  K_2\wedge J -(a_2 - \k b_3 ) (J\wedge P_0 + P_2\wedge K_1 ) \nn\\
&&\back + \k c_3  J\wedge P_1 + \k b_4  K_2\wedge K_1 + \k a_3  P_1\wedge K_1 +  c_2  K_2\wedge P_0 + b_2 K_2\wedge P_2 +  a_5  P_0\wedge P_1 + a_1 P_1\wedge P_2 ,
\nn\\[0.1cm]
&&\back\mback  \delta(P_2)=  \k  a_5  J\wedge K_1 + (a_6   - c_1 ) (\k  J\wedge K_2+ P_0\wedge P_1 )+ (b_2 + \k a_3 ) (  P_0\wedge J +P_1\wedge K_2)\nn \\ 
&&\back + \k    c_3  J\wedge P_2+ \k a_4  K_1\wedge K_2 +  c_2   P_0\wedge K_1 + a_2 P_1\wedge K_1 +   b_5   P_0\wedge P_2  + b_1 P_1\wedge P_2 + \k  b_3  P_2\wedge K_2 ,
\nn\\[0.1cm]
&&\back\mback  \delta(K_1)=  c_2  J\wedge K_1 + (a_1 + b_4) ( J\wedge P_0 + P_1\wedge K_2 ) + (a_6+ c_1) (J\wedge P_1+ P_0\wedge K_2 ) +  b_5   J\wedge P_2 \nn\\
&&\back + a_2 K_1\wedge K_2 +  a_5  P_0\wedge K_1 + c_3   P_0\wedge P_2+ a_4 P_1\wedge K_1+ b_3 P_1\wedge P_2 + b_1 P_2\wedge K_2  ,
\nn\\[0.1cm]\
&&\back \mback \delta(K_2)=  c_2   J\wedge K_2 + b_2 K_1\wedge K_2 + a_1 K_1\wedge P_1 + (a_4 -b_1)(  P_0\wedge J + P_2\wedge K_1 ) + b_5 P_0\wedge K_2  \nn \\
&&\back  +  ( b_6  - c_1) (P_0\wedge K_1+P_2\wedge J )+ a_5   P_1\wedge J +  c_3  P_1\wedge P_0 + b_4 P_2\wedge K_2 +a_3 P_2\wedge P_1,  \label{cocogen} 
\end{eqnarray}
where all the parameters appearing in these expressions must satisfy the constraints (\ref{constraints}).
Obviously, some of the parameters could be shown to be inessential through the appropriate automorphisms, and physical restrictions on the parameters should be also imposed in order to obtain some tractable deformations. Anyhow, these expressions give a clear idea about the size of the zoo of possible quantum deformations of the \adsw algebra. Note also that the dual of $\delta$ (\ref{cocogen}) would provide the most generic first-order \adsw noncommutative spacetime.


\section{Beyond $\kappa$-deformation in (2+1) dimensions}

The generalizations of the $\kappa$-deformation that are contained in the previous characterization of all the quantum \adsw deformations can be identified by imposing that, like in (\ref{cok}), 
\be
\delta(J)=0,\qquad  \delta(P_0)=0.\nn
\ee
These conditions imply that the deformation parameters have to fulfil the relations 
\be
a_i=b_i=0 \quad (i=1,2,3,4);\quad a_5=b_5; \quad a_6=b_6=0; \quad c_2=\k c_3 . \nn
\ee
Under these conditions the resultant $r$-matrix only depends  on {\em three} parameters
\be
r= b_5 (P_1 \wedge K_1+P_2\wedge K_2) + c_1 J\wedge P_0 + c_3(\k  K_1\wedge K_2 + P_1\wedge P_2) ,
\label{rkgen}
\ee
where the $\kappa$-\adsw Planck length parameter is $b_5= -z$ (see (\ref{ca})). These three parameters are free ones, since (\ref{rkgen}) is already a solution of the mCYBE. Moreover,  the Schouten bracket for the $r$-matrix (\ref{rkgen}) reads
\begin{eqnarray}
[[r,r]]=  -2 b_5 c_3 (P_1   \wedge  P_2 \wedge  P_0 + \k  P_0 \wedge K_1 \wedge  K_2 + \k P_1 \wedge  K_1 \wedge  J+\k  P_2 \wedge  K_2 \wedge J)\nonumber\\
 -(    b_5^2  + \k  c_3^2 )(P_1\wedge K_1\wedge P_0+P_1\wedge P_2\wedge J+\k  K_1\wedge K_2\wedge J+ P_0\wedge P_2\wedge K_2),\nn
\end{eqnarray}
and this implies that if  $\k \neq 0$, non-standard (twist) deformations are obtained when 
$b_5=c_3=0$ ({\em i.e.}, the term $J\wedge P_0$ generates a Reshetikhin twist). In the Poincar\'e case ($\k = 0$), the $P_1\wedge P_2$ term provides a second twist.

The full cocommutator that corresponds to the three-parametric $r$-matrix (\ref{rkgen}) is
\begin{eqnarray}
&& \back \delta(J)=\delta(P_0) =0 ,
\nn\\
&&
\back \delta (P_1)= \, c_1  ( P_0\wedge P_2 +\k  K_1\wedge J) - z (\k  K_2\wedge J +P_0\wedge P_1 )+  \k  c_3(J\wedge P_1 + K_2\wedge P_0) ,
\nn\\
&&
\back \delta(P_2)=-   \, c_1  (\k  J\wedge K_2+P_0\wedge P_1 )  -z  (\k  J\wedge K_1+P_0\wedge P_2)    + \k  c_3  (J\wedge P_2 + P_0\wedge K_1) ,
\label{cocogen2} \nn\\
&&
\back \delta(K_1)= \,  c_1 (J\wedge P_1+ P_0\wedge K_2 ) -z (J\wedge P_2 +P_0\wedge K_1) + c_3 (P_0\wedge P_2+\k  J\wedge K_1 ) ,
\nn\\
&&
\back \delta(K_2)=  - \,  c_1 (P_0\wedge K_1+P_2\wedge J ) -z  (P_0\wedge K_2 +P_1\wedge J) + c_3 (P_1\wedge P_0 +\k  J\wedge K_2 ),
\nn
\end{eqnarray}
which leads to the first-order noncommutative spacetime given by
\bea
&& [\hat x_0, \hat x_1]=-z \hat x_1 - c_1 \hat x_2 - c_3\hat\xi_2 ,\nn\\
&& [\hat x_0, \hat x_2]=-z \hat x_2 + c_1 \hat x_1+ c_3\hat\xi_1, \nn \\
&& [\hat x_1, \hat x_2]=0 \nonumber,
\nonumber
\eea
thus providing a two-parametric generalization of the $\kappa$-Minkowski spacetime (\ref{kminkow}). 

To the best of our knowledge, the full quantum \adsw algebra generated by the $r$-matrix (\ref{rkgen}) is not known yet and constitutes and interesting open problem. As a first step, the Poisson counterpart of the corresponding all-orders noncommutative spacetimes could be obtained by computing the PL brackets on the \adsw group through the Sklyanin bracket given by (\ref{rkgen}). Again, the spacetime brackets $\{ x_i,x_j\}$ so obtained will be no longer linear when the curvature $\k$ is turned on.


\section{The (3+1)-dimensional case}

The same Lie bialgebra approach can be applied to study the quantum deformations of the
(3+1) \adsw Lie  algebras, with generators $\{J_i,P_0,P_i, K_i\},\ (i=1,2,3)$ and whose commutation rules are
\be
\begin{array}{lll}
[J_i,J_j]=\varepsilon_{ijk}J_k ,& \quad[J_i,P_j]=\varepsilon_{ijk}P_k  ,&\quad
[J_i,K_j]=\varepsilon_{ijk}K_k ,\\[2pt]\displaystyle{
[P_i,P_j]=-\k\,\varepsilon_{ijk}J_k},&\quad\displaystyle{[P_i,K_j]=-\,\delta_{ij}P_0}
,    &\quad\displaystyle{[K_i,K_j]=-\,\varepsilon_{ijk} J_k}
,\\[6pt][P_0,P_i]=\k  K_i ,&\quad [P_0,K_i]=-P_i  ,&\quad[P_0,J_i]=0 ,
\end{array}
\label{3mas1}
\ee
where $i,j= 1,2,3$, $\varepsilon_{123}=1$ and the $\k$ parameter is again the constant sectional curvature of the associated spacetime such that  $\k=-\Lambda$. Therefore, when $\k>0$ we have the (3+1) AdS Lie algebra $so(3,2)$, when $\k<0$ we have the (3+1) dS Lie algebra $so(4,1)$, and the flat limit $\k=0$ gives rise to the (3+1) Poincar\'e Lie algebra $iso(3,1)$.

The description of the quantum deformations of the Lie algebra (\ref{3mas1}) starts by considering the most
generic $r$-matrix, $r=r^{ij}Y_i\wedge Y_j$ that in this case would depend on 45 deformation parameters
$r^{ij}$ (we recall that again in the (3+1) case all \adsw Lie bialgebras are coboundaries). In particular, if we are interested in generalizations of the $\kappa$-deformation for the \adsw algebra, 
the minimal constraint that we have to impose is that one deformation parameter has to play the role of a Planck length or, equivalently, that the coproduct of the quantum  $P_0$ generator has to be primitive ($\Delta(P_0)=P_0\otimes 1 + 1\otimes P_0$). As we know, this means that we have to impose
\be
\delta(P_0)=[P_0\otimes 1 + 1\otimes P_0,r^{ij}Y_i\wedge Y_j]=0, \nn
\ee
and, as a result, this assumption elliminates 30 of the $r^{ij}$ parameters. If, by following the (2+1) $\kappa$-\adsw case, we further impose that $J_3$   also remains primitive under deformation, 
we get a five-parametric candidate for the $r$-matrix
\bea 
&& r=\z_1(K_1\wedge P_1+K_2\wedge P_2)+\z_2(P_1\wedge P_2+\k K_1\wedge
K_2)\nonumber\\
   &&\qquad \quad +\z_3 P_0\wedge J_3 + \z_4 K_3\wedge P_3+ \z_5 J_1\wedge J_2 
,\label{cb}\nn
\eea 
although now the mCYBE is not automatically fulfilled and leads to the following relations among the quantum deformation parameters:
\bea
&&  \z_1\z_2=0 ,\qquad (\z_1-\z_4)\z_5- {\k} \z_2\z_4=0 ,\nonumber\\
&&\z_2\z_5+  \z_1(\z_1-\z_4)+\k\z_2^2   =0 , \nonumber\\&&
\z_2\z_5-  \z_4(\z_1-\z_4)=0   , \nn \\ && \z_5^2+
{\k}\left(\z_2\z_5- \z_1\z_4    \right)=0. \nonumber 
\eea

Finally, the solution of these equations  leads  to  two  disjoint families of  two-parametric \adsw Lie bialgebras, that are generated by the two following classical $r$-matrices~\cite{prague}:
     \bea&&\!\!\!\!\!\!\!\!\!\!\!\! r_{\z_1,\z_3}=\z_1\left(K_1\wedge
P_1+K_2\wedge P_2+K_3\wedge P_3 \pm  \sqrt{\k}\,  J_1\wedge
J_2\right) +\z_3 P_0\wedge J_3    ,\label{r1}\\ &&\!\!\!\!\!\!\!\!\!\!\!\!
r_{\z_2,\z_3}= \z_2\left(P_1\wedge P_2+\k K_1\wedge K_2-\k\,
J_1\wedge J_2\pm\sqrt{\k}  P_3\wedge K_3\right)  +\z_3 P_0\wedge J_3   ,
\label{r2}
\eea
whose Schouten brackets are $(l=1,2)$:
\bea
&&\!\!\!\!\!\!\!\!\!\!\!\!\!\!\!\!\!\! [[ r_{\z_l,\z_3} , r_{\z_l,\z_3} ]]=A_l
\left({\k}\,J_1\wedge J_2\wedge J_3-\frac 12\varepsilon_{ijk}\left( \k
J_i\wedge K_j\wedge K_k +J_i\wedge P_j\wedge P_k \right ) \right)\cr&&\qquad
\qquad  +A_l \sum_{i=1}^3K_i\wedge P_i\wedge P_0 ,\qquad {\rm where}\quad
A_1={\z^2_1} ,\quad A_2={\z^2_2\k} .
\nonumber
\eea 

Some comments on these two $r$-matrices are pertinent:
\begin{itemize}

\item The $r_{\z_1,0}$ matrix generates the $\kappa$-\adsw deformation in (3+1) dimensions, but the explicit form of its all-orders quantization when $\k\neq 0$ is still an open problem.

\item When $\k=0$, the $r_{\z_1,\z_3}$ matrix generates the twisted $\kappa$-deformation of the (3+1) Poincar\'e algebra whose full expressions were given in~\cite{Dasz}.

\item If we compare the (3+1) $r$-matrix (\ref{r1}) with the (2+1) $r$-matrix (\ref{rkgen}), we realize that  the $c_3$ and the $b_5$ terms of the latter are not compatible with the $\kappa$-deformation in (3+1) dimensions, and can only be simultaneously considered in the deformation of the second type generated by (\ref{r2}). The full quantization of the latter is also unknown yet.

\end{itemize}

The first-order noncommutative \adsw spaces associated to these two families of Lie bialgebras are deduced from the dual cocommutators, and are given by the following commutation rules involving the (3+1) quantum coordinates 
   $\{ \hat\theta, \hat
x_\mu,  \hat\xi_i \} $     dual to  $\{J,P_\mu,K_i\}$ $(\mu=0,1,2,3)$
\bea
r_{\z_1,\z_3}:&& [\hat x_0,\hat x_i]=-{\z_1}\,\hat x_i 
+\z_3\,\varepsilon_{ij3}\, \hat x_j  ,\qquad [\hat x_i,\hat
x_j]=0. \nn  \\
r_{\z_2,\z_3}:
&& [\hat x_0,\hat
x_I]=-{\z_2}\,\varepsilon_{IJ3}\,\hat\xi_J   
+\z_3\,\varepsilon_{IJ3}\, \hat x_J  , \quad   [\hat x_0,\hat
x_3]=\pm{\z_2}\,\sqrt{\k}\,\hat x_3 , \nonumber\\
&&    [\hat x_I,\hat
x_3]= \z_2 \hat\theta_I ,\qquad  [\hat x_1,\hat x_2]=0  ,\qquad I,J=1,2. \nn
\eea
Note that the first deformation gives rise to a generalization of the $\kappa$-Minkowski space, that is recovered when $\z_3=0$. Again, no dependence on the curvature $\k$ appears in this first-order quantum group, although a nonlinearity ruled by the cosmological constant is expected when the full quantum group is constructed.
On the contrary, the quantum spacetimes coming from the second deformation have a completely different shape:  the deformation linked to $\z_2$ introduces the quantum rotation and boost coordinates into the spacetime commutation rules, that are no longer a subalgebra and, moreover, the cosmological constant is already manifest at the first-order of these quantum group relations. 

Finally, we would like to recall that one of the main features of the (3+1)  $\kappa$-Poincar\'e algebra is the fact that the rotation subalgebra remains undeformed. Therefore, the isotropy of the `quantum space' is preserved, since the $\kappa$-Minkowski spacetime has the same type of commutation rules for the three quantum space coordinates. This seems to be a quite reasonable physical condition and, apparently, the generalization of the $\kappa$-Poincar\'e deformation provided by $r_{\z_1,\z_3}$ would break this `quantum isotropy' property. However, we would like to stress that if we consider the
following automorphism  of the \adsw algebra
\be
\begin{array}{ll}\displaystyle{\widetilde
Y_1=-\frac{2}{\sqrt{6}}\,Y_2+\frac{1}{\sqrt{3}}\,Y_3},&\qquad\displaystyle{Y_1=
\frac{1}{\sqrt{2}}\left( \widetilde  Y_2- \widetilde
Y_3\right)},\\[6pt]\displaystyle{\widetilde
Y_2=\frac{1}{\sqrt{2}}\,Y_1+\frac{1}{\sqrt{6}}\,Y_2+\frac{1}{\sqrt{3}}\,Y_3},&\qquad\displaystyle{Y_2=
\frac{1}{\sqrt{6}}\left(-2 \widetilde  Y_1+ \widetilde Y_2+ \widetilde
Y_3\right)},\\[6pt]\displaystyle{\widetilde
Y_3=-\frac{1}{\sqrt{2}}\,Y_1+\frac{1}{\sqrt{6}}\,Y_2+\frac{1}{\sqrt{3}}\,Y_3},&\qquad\displaystyle{Y_3=
\frac{1}{\sqrt{3}}\left(  \widetilde  Y_1+ \widetilde Y_2+ \widetilde
Y_3\right)}, \\[6pt]\mbox{for}\quad Y_i\in\{J_i,P_i,K_i\},&\qquad \widetilde
P_0=P_0 ,
\end{array}
\label{ma}\nn
\ee
where $\tilde Y_i$ denote the transformed generators, and if we take $
\tilde z_3=  z_3/\sqrt{3}
$, we get
\bea
&&\!\!\!\!\!\!\!\!\!\!\!\!\!\!\!\! r_{\z_1,\tilde\z_3}=\z_1\left(\tilde
K_1\wedge \tilde P_1+\tilde K_2\wedge\tilde P_2+\tilde K_3\wedge \tilde P_3 \pm 
\frac{\sqrt{\k}}{\sqrt{3}}\,\left( \tilde J_1\wedge \tilde J_2+ \tilde
J_2\wedge \tilde J_3+\tilde J_3\wedge\tilde  J_1\right)\right)\cr  &&
\qquad\qquad +\tilde\z_3\tilde P_0\wedge \left(\tilde J_1+\tilde J_2+\tilde J_3 
\right)  .
 \label{mb}\nn
\eea
This means that in this new \adsw basis, the `broken' isotropy of the $r_{\z_1,\z_3}$ deformation  can be restored, since the role of the three rotation generators is exactly the same. This indicates that some new multiparametric deformations of the (A)dS and Poincar\'e algebras could be considered as reasonable symmetries on a physical basis.

Summarizing, in this work we have shown that the classification of the quantum deformations of the kinematical symmetries of (2+1) and (3+1) relativistic spacetimes can be approached in a very efficient and computationally tractable way by analyzing the first-order deformation given by the corresponding Lie bialgebra structures. We think that this viewpoint could be useful in the near future in order to explore the role that multiparametric quantum groups with non-vanishing cosmological constant could play in a quantum gravity context.



\section*{Acknowledgments}

This work was partially supported by the Spanish MICINN under grant MTM2010-18556 (with EU-FEDER support).


\end{document}